\documentclass[12pt]{article}
\usepackage{amssymb}
\usepackage{amscd}
\usepackage{amsmath}

\begin{document}
\begin{titlepage}

{\hbox to\hsize{\hfill August 2015 }}

\bigskip \vspace{3\baselineskip}

\begin{center}
{\bf \large 
The Effective MSSM }

\bigskip

\bigskip

{\bf Archil Kobakhidze and Matthew Talia \\ }

\smallskip

{ \small \it
ARC Centre of Excellence for Particle Physics at the Terascale, \\
School of Physics, The University of Sydney, NSW 2006, Australia \\
E-mail: archilk@physics.usyd.edu.au, mtal0627@uni.sydney.edu.au
\\}

\bigskip
 
\bigskip

\bigskip

{\large \bf Abstract}

\end{center}
\noindent 
We suggest an effective field theory framework to discuss deviations from the minimal supersymmetric Standard Model (MSSM) which is based on an alternative arrangement of the gauge-Higgs sector. In this effective MSSM (EffMSSM)  nonlinearly realised $SU(2)\times U(1)$ gauge sector is described by an $SU(2)\times U(1)$-valued massive vector superfield, which contains a neutral CP-even and charged Higgs fields, while another neutral CP-even Higgs and the neutral CP-odd Higgs fields are residing in an $SU(2)\times U(1)$-singlet chiral superfield. Although the new theory contains the same particle content as the conventional MSSM, the unconventional representation of superfields allows for new type of interactions, which may lead to a significant modification of the phenomenology. As an illustrative example we consider EffMSSM with modified Higgs and electroweak gauge sector augmented by gaugino soft supersymmetry breaking masses, $M_i~ (i=1,2,3)$ and the Standard Higgs soft-breaking masses, $m_{H_u}=m_{H_d}$ and $B_{\mu}$, and point out distinct features in the Higgs and gaugino sectors as compared to MSSM. In particular, we show that the lightest neutral CP-even Higgs boson with mass $\sim 125$ GeV can be easily accommodated within EffMSSM.    

\end{titlepage}
\baselineskip=16pt
\section{Introduction}

The null results of LHC searches for supersymmetric (SUSY) particles during Run I have significantly constrained the simplest supersymmetric models of particle physics and  the minimal supersymmetric Standard Model (MSSM) in particular. Furthermore, the discovery of the Higgs-like particle with mass $m_h\approx 125$ GeV, which in MSSM is associated with the lightest CP-even Higgs boson, essentially excludes natural versions of MSSM, except the case of compressed sparticle sprectrum \cite{Han:2013usa}. Indeed, 125 GeV Higgs boson can be accommodated within MSSM providing stops are sufficiently heavy, but then substantial tuning of parameters is required to obtain the correct masses for the electroweak gauge bosons. Furthermore, Higgs coupling measurements already started to constraint scenarios with relatively light stops and large stop-Higgs trilinear coupling \cite{Arvanitaki:2011ck}. Nevertheless, at this stage it is still premature to attribute this problem to SUSY. In fact, the problem of Higgs mass and naturalness is specific for MSSM and may be avoided in extended theories, such as the next-to-minimal  supersymmetric Standard Model (NMSSM) \cite{Gherghetta:2012gb} or SUSY models with an extra gauge symmetry \cite{Batra:2003nj}.

In this paper we would like to suggest a framework within which deviations from the MSSM can be conveniently parametrised and thus possible deviations from the minimal model can be discussed in a model-independent way.  We dubbed this framework effective MSSM (EffMSSM).  It is based on an alternative arrangement of superfield representations in the gauge-Higgs sector within the $SU(2)\times U(1)$ electroweak symmetry being nonlinearly realised\footnote{Similar approach has been taken to analyse deviations from the Standard Model Higgs sector in \cite{Espinosa:2012ir, Kobakhidze:2012wb} }. Namely,  the $SU(2)\times U(1)$ gauge sector  is described by an $SU(2)\times U(1)$-valued massive vector superfield, beside massive electroweak gauge fields and corresponding gauginos, contains also $H^0, H^{\pm}$ Higgs fields. In addition, we introduce $SU(2)\times U(1)$-singlet chiral superfield where $h^0$ CP-even and $A^0$ CP-odd Higgs fields are residing. Thus, the particle content is exactly the same as in the MSSM. However, due to the unconventional superfield representation many new types of interactions become possible. MSSM is attained as a particular case of EffMSSM. A similar model with different emphases has been considered previously in \cite{Ferrara:1992sd}.  

We note that existing large uncertainties in Higgs couplings measurements precludes from the definite conclusion on the nature of the electroweak symmetry. If nonlinearly realised, the electroweak gauge theory becomes strongly interacting at high energies, the famous example being $WW\to WW$ scattering in the Higgsless Standard Model (SM). It is expected that at high energies new resonances show up, which unitarise rapid, power-law  growth of scattering amplitudes with energy in perturbation theory. However, the scale where new physics is expected to emerge crucially depends on the specific process considered. For example, in the SM with anomalous top-Yukawa couplings perturbative unitarity is violated in the $t\bar t \to WW$ at energies $\sim 10$ TeV. New physics at such high energies may escape the detection at LHC, and, in situations like these, precision measurements of deviations from SM physics parametrized within the effective theories based on nonlinear realisation become imperative.      

In the next section we set up the EffMSSM framework. In section \ref{EWSB} we discuss the electroweak symmetry breaking within MSSM and calculate the mass spectrum in the following section. We conclude in section \ref{conc}.      

\section{Description of the model} 
We describe the broken phase of the electroweak $SU(2)\times U(1)_Y$ gauge theory with an residual unbroken $U(1)_{EM}$ in a model-independent way by introducing nonlinear chiral superfield
\begin{equation}
	U=e^{\frac{i}{2}\xi_i\sigma_i}~,~~{\rm det} U=1,
	\label{1}
\end{equation}  
where $\sigma_{i}~(i=1,2,3)$ are the three Pauli matrices. The chiral superfields $\xi_i$ contain the three Goldstone bosons spanning $SU(2)\times U(1)_Y/U(1)_{EM}$ coset space and their SUSY counterparts, which are pseudo-Goldstone bosons. The first three  represent longitudinal degrees of freedom of the electroweak $W$ and $Z$ vector particles, while the pseudo-Goldstones complete the electroweak massive vector supermultiplets.  The superfield $U$ transforms as the following under the electroweak gauge group:
	\begin{equation}
	U\rightarrow e^{\frac{i}{2}\Lambda_{i}\sigma_{i}}Ue^{-\frac{i}{2}\Sigma\sigma_{3}}~,
\label{2}	
	\end{equation}
where $\Lambda_{i}$ and $\Sigma$ are chiral superfields for $SU(2)$ and $U(1)_Y$ supergauge transformation parameters, respectively.  In addition, we introduce  a singlet chiral superfield $S$, such that the two Higgs superfields of the conventional MSSM, $H_u$ and $H_d$, can be identified with the composite superfield $SU$ as follows: 
	\begin{equation}
	\Phi\equiv SU=
	\begin{pmatrix} H_{u}^{0}  &  H_{d}^{-}\\
	H_{u}^{+} & H_{d}^{0}
	\end{pmatrix}
	\label{3}
	\end{equation}
It is easy to see, that	
		\begin{equation}
	{\rm det}\Phi= S^2 = H_u H_d~,
	\label{4}
	\end{equation}
and 
\begin{eqnarray}
H_{u(d)}^0=S\cos\left(\frac{\sqrt{\xi_i\xi_i}}{2}\right) \pm iS\frac{\xi_3}{\sqrt{\xi_i\xi_i}}\sin\left(\frac{\sqrt{\xi_i\xi_i}}{2}\right)~, 
\label{5}
 \\ \nonumber
\\
H_{u(d)}^{+(-)}=iS\frac{\xi_{\pm}}{\sqrt{\xi_i\xi_i}}\sin\left(\frac{\sqrt{\xi_i\xi_i}}{2}\right)~,
\label{6} 
\end{eqnarray}
$\xi_{\pm}=\frac{1}{\sqrt{2}}(\xi_1\pm i\xi_2)$~.
	
The most general, renormalizable Lagrangian for the gauge-Higgs sector comprises then of the following D-terms:
	\begin{eqnarray}
	\mathcal{L}_{\rm HG}	&=&	\left[Tr\left(\Phi^{\dagger}e^{W}\Phi e^{B}\right)\right]_{D}+\kappa^{2}\left[Tr\left(U^{\dagger}e^{W}Ue^{B}\right)\right]_{D} \nonumber \\ 
	&&+\left[\alpha Tr\left(\Phi^{\dagger}e^{W}Ue^{B}\right)+\alpha^{*}Tr\left(U^{\dagger}e^{W}\Phi e^{B}\right)\right]_{D} +\beta \left[\bar{S}S\right]_{D}~,
	 \label{7}
	\end{eqnarray}
where $W=gW_i\sigma_i$ and $B=g'Y\sigma_{3}$ are, respectively, $SU(2)$ and $U(1)_Y$ gauge superfields in the adjoint representation of the electroweak gauge group:	\begin{equation}
	e^{W}\rightarrow e^{+i\Lambda^{\dagger}}e^{W}e^{-i\Lambda},\quad e^{B}\rightarrow e^{i/2\Sigma\sigma_{3}}e^{B}e^{-i/2\Sigma^{\dagger}\sigma_{3}}~.
	\label{8}
	\end{equation}
The extra parameters $\kappa$ and $\alpha$ have mass dimension one, while $\beta$ is dimensionless. They parametrise deviations from the MSSM gauge-Higgs sector, which is represented by the first term on the rhs of (\ref{7}). 

The Higgs-Yukawa sector of the theory is described by the F-term of the following superpotential:
\begin{equation}
	W_{HY}=\bar{u}\left(\mathbf{y_{u}}\Phi +\mathbf{y'_{u}}U\right)\chi_u Q 
	-\bar{d}\left(\mathbf{y_{d}}\Phi+\mathbf{y'_{d}}U\right)\chi_dQ
	-\bar{e}\left(\mathbf{y_{e}}\Phi+\mathbf{y'_{e}}U\right)\chi_d L~, 
	\label{9}
	\end{equation}
where $\chi_u=\left(1\,\,0\right)^{\rm T}$ and $\chi_d=\left(0\,\,1\right)^{\rm T}$, and $\bar u$, $\bar d$, $\bar e$, $Q$ and $L$ are quark and lepton chiral superfields, all 3-vectors in the flavour space. The 3-by-3 matrices $\mathbf{y_{u,d,e}}$ with dimensionless entries are the conventional Higgs-Yukawa couplings, while $\mathbf{y'_{u,d,e}}$ are extra non-linear mass matrices. The $\mathbf{y'_{u,d,e}}\to 0$ limit reproduces the MSSM Higgs-Yukawa superpotential.    

Finally, the renormalizable Higgs superpotential involving the superfield $S$ takes the form:
\begin{equation}
W_{\rm H}= \frac{\lambda}{3}S^{3}+\frac{\mu}{2}S^{2}-\tau S~.
\label{10}
\end{equation}
Note that, because of the relation (\ref{4}), the quadratic $S^2$ term in the above equation reproduces (up to a normalisation related factor 1/2, see below) the usual MSSM $\mu$-term. The dimensionless cubic coupling $\lambda$ and the $\tau$ parameter of (mass)$^2$ dimension parametrise deviations from the MSSM Higgs superpotential.

Augmented by the SUSY soft breaking terms, Eqs. (\ref{7}), (\ref{9}) and (\ref{10}), describes the general EffMSSM, which involves new interactions, while having  the same particle content as the MSSM. As a result, the model becomes significantly more complicated and phenomenologically richer, but also more flexible to accommodate experimental constraints. However, our aim here does not include a comprehensive study of phenomenological consequences of EffMSSM. Instead, in what follows, we concentrate on the electroweak symmetry breaking and the related mass spectrum within a simplified version of the general model, which is realistic and, at the same time, tractable analytically.        

\section{Electroweak symmetry breaking in EffMSSM}   \label{EWSB}
To avoid notational clutter, in this section we denote the scalar components by the same letters as their corresponding chiral superfields, e.g. $\left. S \right \vert_{\theta=0}=S$,  etc. Also, for phenomenological reasons we assume that squarks and sleptons do not develop vacuum expectation values, and, hence, for the purpose of this section we concentrate on terms for the scalar potential stemming  from Eqs. (\ref{7}) and (\ref{10}). 
To further simplify the matter, we can use the gauge freedom to remove the electrically neutral would be Goldstone boson from the scalar potential by setting ${\rm Re (\xi_3)}=0$.  

As usually, supersymmetry at low energies is broken by a set of soft breaking parameters. For discussion of the electroweak symmetry breaking in the tree-level approximation it is sufficient to consider soft scalar masses for $S$ and $\xi_i$ fields:
\begin{equation}
V_{\rm soft}= \left(\frac{1}{2}m^2_S S^2 + {\rm h.c.}\right)+\frac{A}{2}{\rm Tr}\left(\Phi^{\dagger}\Phi\right)+\frac{B}{2}{\rm Tr}\left(\Phi^{\dagger}\Phi\sigma_3\right)
\label{11}
\end{equation}
 In the MSSM limit, the above soft breaking masses correspond to:
 \begin{equation}
 A=m_{H_u}^2+m_{H_d}^2~,~~B=m_{H_u}^2-m_{H_d}^2~,~~m_S^2= 4B_{\mu}~. 
 \label{11}
 \end{equation} 
We note that other gauge invariant soft breaking terms in the Higgs sector can also be introduced within the EffMSSM such as, $S^*S,~{\rm Tr}\left(U^{\dagger}U\right),~{\rm Tr}\left(U^{\dagger}U\sigma_3\right)$. The relation to the MSSM soft parameters become more complicated, and we do not consider them here.  

\begin{equation}
	V_{\rm H}=	\left|\lambda S^{2}+\mu S-\tau\right|^{2} + \left(S\bar{S}+\alpha\bar{S}+\alpha^{*}S+\kappa^{2}\right)^{2}V_{D}+V_{\rm soft}~,	
\label{12}	
	\end{equation}
	where the D-term contribution can be written in exact form as:
	\begin{eqnarray}
V_{D}	&=&	\frac{g^{2}+g'^{2}}{2}\left[\frac{i\xi_{3}}{\sqrt{\xi_{i}\xi_{i}}}\cos\left(\frac{\sqrt{\xi^*_{i}\xi^*_{i}}}{2}\right)\sin\left(\frac{\sqrt{\xi_{i}\xi_{i}}}{2}\right)-\frac{\bar{i\xi_{3}}}{\sqrt{\xi^*_{i}\xi^*_{i}}}\cos\left(\frac{\sqrt{\xi_{i}\xi_{i}}}{2}\right)\sin\left(\frac{\sqrt{\xi^*_{i}\xi^*_{i}}}{2}\right)\right. \nonumber \\
		&&\left.+\frac{\xi^*_{+}\xi_{+}-\xi^*_{-}\xi_{-}}{\sqrt{\xi_{i}\xi_{i}\xi^*_{j}\xi^*_{j}}}\sin\left(\frac{\sqrt{\xi_{i}\xi_{i}}}{2}\right)\sin\left(\frac{\sqrt{\xi^*_{i}\xi^*_{i}}}{2}\right)\right]^{2} \nonumber \\
		&&+g'^{2}\left|\frac{i\xi_{+}}{\sqrt{\xi_{i}\xi_{i}}}\sin\left(\frac{\sqrt{\xi_{i}\xi_{i}}}{2}\right)\cos\left(\frac{\sqrt{\xi^*_{i}\xi^*_{i}}}{2}\right)-\frac{i\bar{\xi_{-}}}{\sqrt{\xi^*_{i}\xi^*_{i}}}\sin\left(\frac{\sqrt{\xi^*_{i}\xi^*_{i}}}{2}\right)\cos\left(\frac{\sqrt{\xi_{i}\xi_{i}}}{2}\right)\right. \nonumber \\
		&&\left.+\frac{\xi^*_{-}\xi_{3}-\xi_{+}\xi^*_{3}}{\sqrt{\xi_{i}\xi_{i}\xi^*_{j}\xi^*_{j}}}\sin\left(\frac{\sqrt{\xi_{i}\xi_{i}}}{2}\right)\sin\left(\frac{\sqrt{\xi^*_{i}\xi^*_{i}}}{2}\right)\right|^{2}
	\end{eqnarray} 
As in the MSSM, the vanishing charged fields minimize the potential, so $\xi_+=\xi_-=0$ in the vacuum. Hence, Eq. (\ref{12}) takes the simpler form:
	\begin{eqnarray}
	V_{\rm H}	&=&	\left|\lambda S^{2}+\mu S-\tau\right|^{2} \nonumber \\
	&&+\frac{g^2+g'^2}{2}\left(SS^*+\alpha S^*+\alpha^{*}S+\kappa^{2}\right)^{2}\sinh^{2}\xi  \nonumber\\
	&&+\frac{A}{2}SS^*\cosh\xi-\frac{B}{2}SS^*\sinh\xi+\left(\frac{1}{2}m_S^2S^{2}+{\rm h.c.}\right)~,	
\label{13}	
	\end{eqnarray}
where $\xi \equiv{\rm Im}(\xi_3)$. 
Analysing the above potential, first we note that, unlike the MSSM, the electroweak symmetry breaking in EffMSSM can also be achieved in the supersymmetric limit, $A=B=m_S=0$. Indeed, the D- and F-flatness conditions,
\begin{equation}
\xi=0~,~~\lambda S^{2}+\mu S-\tau =0~,
\label{14}
\end{equation}
respectively, lead to a non-zero vacuum expectation value for the singlet field. Using the relations for the vacuum expectations in the linear realization,
	\begin{equation}
	e^{\left\langle \xi\right\rangle }\equiv\tan\beta,\quad\left\langle S\right\rangle ^{2}= v_{u}v_{d}~,
\label{15}	
	\end{equation}
	we then have in this case $v_u=v_d$, that is, $\left\langle S\right\rangle ^{2}=v^2/2$, where $v=\sqrt{v_u^2+v_d^2}$.

Note that, the expectation value $\left\langle S\right\rangle$ can be complex, thus breaking CP spontaneously. To simplify the matter we only consider the case of real $\lambda, \mu$ and $\tau$ parameters. Inspecting the F-flatness condition (\ref{14}), we find that for $\lambda\tau<0$ and $|\mu|<2\sqrt{-\lambda\tau}$, the expectation value is complex with the phase angle $\theta$ and modulus $|v|$ given as: 
 \begin{equation}
 \cos\theta=-\frac{\mu}{2\sqrt{-\lambda\tau}}~,~~|v|=\sqrt{-\frac{2\tau}{\lambda}}~.
 \label{16}
 \end{equation}
If, however, $\lambda\tau >0$, the expectation value is real and for $\theta=0$ there are 2 degenerate solutions for $v$:
\begin{equation}
v=-\frac{\mu}{2\sqrt{2}\lambda}\left(1\pm\sqrt{1+\frac{16\lambda\tau}{\mu^2}}\right)~.
\label{17}
\end{equation}
 For $\theta=\pi$, $v\to -v$, and the above two solutions are simply swapped. Obviously, each of them can be associated with the electroweak vacuum.

The soft breaking masses in (\ref{13}) lift the flatness of the Higgs potential. However, unlike the MSSM case, in the limit of vanishing $B$,  we can still maintain D-flatness, $\xi=0$ [$\tan\beta=1$] for non-zero $\langle S \rangle$. Again, $\langle S \rangle$ can be complex, but we focus on real CP-conserving solutions $\theta=0,\pi$ in what follows. The vacuum expectation value is then a solution to the following extremum equation:
\begin{equation}
2\lambda v^3+\left(2\mu^2+m_S^2-4\lambda\tau\right)v\pm \sqrt{2}\mu\left(3\lambda v^2-2\tau\right)=0~,
\label{18}
\end{equation} 
where $\pm$ signs correspond to $\theta=0$ and $\pi$, respectively. Note that for  $\mu\tau\neq 0$ all the solutions of the above equation are non-trivial.  
 
To compute physical quantities, we need to canonically normalize kinetic terms for physical fields. This can be achieved by the following  
rescaling of $S$ and $\xi_i$ chiral superfields:	
\begin{equation}
	S\rightarrow  \sqrt{2+\beta} S,\quad \xi_{i}\rightarrow \rho \xi_{i}~,
	\end{equation}
	where
	\begin{equation}
	\rho\equiv\frac{v^{2}}{4}+\frac{{\rm Re}(\alpha) v}{\sqrt{2}}+\frac{\kappa^{2}}{2}
	\end{equation}
In this framework, the Z and W bosons are expressed as:
	\begin{equation}
	m_{Z}^{2}=\left(g^{2}+g'^{2}\right)\frac{\Delta}{2},\quad m_{W}^{2}=g^{2}\frac{\Delta}{2}
	\end{equation}
	where
	\begin{equation}
	\Delta = 4\kappa^{2}+\frac{4\sqrt{2}{\rm Re}(\alpha) v}{\sqrt{2+\beta}}+\frac{2v^{2}}{2+\beta}\approx\left(174~{\rm GeV}\right)^2
	\end{equation}
It is easy to see that one can recover the standard expressions for Z and W masses when the non-minimal parameters, $\kappa, \alpha$ and $\beta$  are set to zero, $\Delta=v^2$.  
	
\section{Mass spectrum}
In this section we compute tree-level mass spectrum of the Higgs sector particles of the EffMSSM described above. 	
	\subsection{Scalar Masses}
For the pair   of neutral CP-even scalar fields  $({\rm Re}(S), {\rm Im}(\xi_{3}))$ we find that the mass matrix is diagonal,
	\begin{equation}
\begin{pmatrix}
\frac{4m_{S}^{2}+4\mu^{2}-8\lambda \tau}{2+\beta}+\frac{12\sqrt{2}\lambda\mu v}{\left(2+\beta\right)^{3/2}}+\frac{12\lambda^{2}v^{2}}{\left(2+\beta\right)^{2}} & 0\\
0 & \frac{\left(g^{2}+g'^{2}\right)}{8}\frac{\Delta^{2}}{\rho}+\frac{Av^2}{\rho}
\end{pmatrix}~,
	\end{equation}
and hence there is no mixing between these states. For the MSSM gauge-Higgs sector, $\kappa=\alpha=\beta=0$, the masses of these states read:
\begin{eqnarray}
m_{H^0_1}^2=2\mu^2+2\lambda\left(3\mu v+3\lambda v^2 -2\tau \right) + 2m_S^2+A
\end{eqnarray}
and
	\begin{equation}
	m_{H^0_2}^{2}= m_{Z}^{2}+4A~.
	\end{equation}
The second state $H^0_2$ is a partner of the $Z$ gauge boson within the massive gauge supermultiplet, as it becomes degenerate with $Z$ boson in the limit $A\to 0$. \emph{A priori}, each of this states can be identified with the LHC Higgs boson. E.g., in \cite{Fayet:2014pia} Z-boson partner state $H^0_2$ has been identified with the observed boson. Within the given framework, however,  it would be more natural to identify the first state $H^0_1$ with the observed resonance, as in the limit  $\kappa=\alpha=\beta=0$ its interactions with the electroweak gauge bosons would exactly coincide with those of the Standard Model Higgs.

	The pair of neutral pseudo-scalars $({\rm Im}(S),{\rm Re}(\xi_{3}))$ has the following mass matrix:	
	\begin{equation}
\begin{pmatrix}
\frac{-4m_{s}^{2}+4\mu^{2}+4\lambda\tau}{2+\beta}+
\frac{4\sqrt{2}\lambda\mu v}{\left(2+\beta\right)^{3/2}}+\frac{4\lambda^{2}v^{2}}{\left(2+\beta\right)^{2}}+\frac{2A}{2+\beta} & 0\\
0 & 0
\end{pmatrix}
	\end{equation}
	As has been noted before ${\rm Re}(\xi_3)$ is the neutral would-be Goldstone state 'eaten up' by the $Z$ boson. Another pseudo-scalar state is an equivalent of  $A^0$ of MSSM. For $\kappa=\alpha=\beta=0$ its mass reads:
	\begin{equation}
	m_{A^{0}}^{2}= 2\mu^2+\lambda\left(\lambda v^2+2\mu v +2\tau\right)+A-m_S^2	
	\end{equation}

Finally, two pairs of charged states $\left({\rm Re}(\xi_+),{\rm Re}(\xi_-) \right)$ and $\left({\rm Im}(\xi_+),{\rm Im}(\xi_-) \right)$ have identical degenerate mass matrices:
	\begin{equation}
\begin{pmatrix}\frac{g^{2}}{16}\frac{\Delta^{2}}{\rho}+\frac{Av^2}{2\rho} & -\frac{g^{2}}{16}\frac{\Delta^{2}}{\rho}-\frac{Av^2}{2\rho}\\
-\frac{g^{2}}{16}\frac{\Delta^{2}}{\rho}-\frac{Av^2}{2\rho} & \frac{g^{2}}{16}\frac{\Delta^{2}}{\rho}+\frac{Av^2}{2\rho}
\end{pmatrix}
	\end{equation}
	The massless eigenstates are identified with the longitudinal states of $W^{\pm}$ gauge bosons. The mass of the physical charged Higgs is given by ($\kappa=\alpha=\beta=0$):
	\begin{equation}
	m_{H^{\pm}}^{2}=m_{W}^{2}+4A
	\end{equation}	
We observe that this charged scalars are degenerate with $W^{\pm}$ in the limit $A\to 0$, as they represent supersymmetric partners of the $W^{\pm}$ gauge bosons within the massive gauge supermultiplet. 	
	
	\subsection{Neutralinos and Charginos}
	
In the basis of fermionic eigenstates $(\tilde{B},\tilde{W_0},\tilde{\xi_3},\tilde{S})$ one can compute the mass matrix for the neutralinos in the non-linear framework:
	\begin{equation}
\mathbf{M_{\tilde{N}}}=\begin{pmatrix}M_{1} & 0 & \frac{ig'}{\sqrt{2}v}\Delta & 0\\
0 & M_{2} & -\frac{ig}{\sqrt{2}v}\Delta & 0\\
\frac{ig'}{\sqrt{2}v}\Delta & -\frac{ig}{\sqrt{2}v}\Delta & 0 & 0\\
0 & 0 & 0 & \mu+\sqrt{2}\lambda v
\end{pmatrix}
  \label{neutralinomm}
	\end{equation}
	
First, we note that the singlino  state $\tilde S$ is decoupled from the rest of the neutral fermionic states. Also, in the limit of restored supersymmetry, $M_{1,2}\to 0$, there is one massless neutralino, a supersymmetric partner of the would-be Goldstone states. The two other neutralino states are degenerate with the $Z$ gauge boson. This is of course due to the fact that these states furnish the neutral massive vector  supermultiplet. Supersymmetry breaking removes this degeneracy. Indeed, for the sake of simplicity, let us consider the case $M_1=M_2\equiv M$ and $\Delta/v \ll M$. The neutralino spectrum is then given as:
	\begin{eqnarray}
	m^2_{\tilde{N_{1}}}&\approx&\frac{m_{Z}^{4}\Delta^{2}}{M^{2}v^{4}} \\
	m^2_{\tilde{N_{2}}}&=&\left|\mu+\sqrt{2}\lambda v\right|^{2} \\
	m^2_{\tilde{N_{3}}}&\approx&M^{2}-\frac{m_{Z}^{4}\Delta^{2}}{M^{2}v^{4}}\\
	m^2_{\tilde{N_{4}}}&=& M^2 	
	\end{eqnarray}
Hence, the LSP in this framework can be quite light and the two heaviest states are nearly degenerate, $m^2_{\tilde{N_{4}}}-m^2_{\tilde{N_{3}}}\approx m^2_{\tilde{N_{1}}}$. This may imply interesting phenomenological consequences for dark matter.   	
	
	For the chargino states  $(\tilde{W_+},\tilde{\xi_+},\tilde{W_-},\tilde{\xi_-})$ we obtain the following 4 x 4 symmetric mass matrix:
	\begin{equation}
	\mathbf{M_{\tilde{C}}}=\begin{pmatrix}0 & \mathbf{C}\\
	\mathbf{C} & 0
	\end{pmatrix} \label{charginomat}
	\end{equation}
	
	\begin{equation}
	\mathbf{C}=\begin{pmatrix}M_{2} & -\frac{2ig}{v}\Delta\\
-\frac{2ig}{v}\Delta & 0
\end{pmatrix}
	\end{equation}
	The doubly degenerate eigenvalues of the 4 x 4 matrix are:	
	
	\begin{equation}
	m_{\tilde{C}_{1},}^{2}m_{\tilde{C}_{2},}^{2}=\frac{M_{2}^{2}}{2}+\frac{4g^{2}\Delta^{2}}{v^{2}}\mp\sqrt{\frac{4g^{2}M_{2}^{2}\Delta^{2}}{v^{2}}+\frac{M_{2}^{4}}{4}}
	\end{equation}
	where as usual the subscripts order increasingly heavy states. In the supersymmetric limit $M_2\to 0$ one chargino is massless and represent the supersymmetric partner of the charged would-be Goldstone boson, while another is degenerate with $W^{\pm}$ gauge bosons. We see again that these state furnish  massive charged vector supermultiplet. Assuming again $\Delta/v \ll M$ the above masses reduce to:
	\begin{eqnarray}
		m^2_{\tilde{C_{1}}}&\approx &\frac{64m_{W}^{4}\Delta^{2}}{M_{2}^{2}v^{4}}
   \\
		m^2_{\tilde{C_{2}}}&\approx &M_{2}^{2}+\frac{64m_{W}^{4}\Delta^{2}}{M_{2}^{2}v^{4}}
	\end{eqnarray}
Hence, one chargino eigenstate can be relatively light.  

\section{Conclusion} \label{conc}
In this paper we have constructed a supersymmetric extension of the Standard Model, which is based on the non-linear realisation of $SU(2)\times U(1)_Y$ supergauge symmetry, the EffMSSM. The gauge-Higgs sector of EffMSSM comprises of massive electroweak vector superfields and the electroweak singlet chiral superfield and no new states have been introduced. We also have established the relation between EffMSSM and MSSM, the later being a particular case of the former one.

Despite the fact the particle content of EffMSSM is the same as in MSSM, non-linearly realised electroweak gauge invariance allows new interactions, which significantly impact the phenomenology of the model. In particular, the electroweak symmetry breaking exhibits several new features, such as the possibility to develop a non-zero electroweak vacuum expectation value in the supersymmetric limit and along the D-flat direction when supersymmetry is broken. The mass spectrum of sparticles and Higgs bosons is altered correspondingly, and the 125 GeV LHC resonance can be comfortably accommodated. 

Since the nature of the electroweak symmetry breaking is not fully understood yet, one should explore wider possibilities beyond the simplest MSSM. The approach described in this paper provides a model-independent framework for such phenomenological studies.           

\paragraph{Acknowledgements.} This work was partially supported by the Australian Research Council. AK was also supported in part by the Rustaveli National Science Foundation under the projects No. DI/8/6-100/12 and No. DI/12/6-200/13.

\end{document}